\documentclass[aps,prl,twocolumn,showpacs,preprintnumbers,amsmath,amssymb]{revtex4}

\usepackage[english]{babel}
\usepackage{graphicx}
\usepackage{subfigure}
\usepackage{dcolumn}
\usepackage{bm}
\usepackage{psfrag}

\begin{document}

\preprint{}

\title{Path Integral Ground State study of 2D solid $^4$He}
\author{E. Vitali}
 \affiliation{Dipartimento di Fisica, Universit\`a degli Studi
              di Milano, via Celoria 16, 20133 Milano, Italy}
\author{M. Rossi}
 \affiliation{Dipartimento di Fisica, Universit\`a degli Studi
              di Milano, via Celoria 16, 20133 Milano, Italy}
\author{F. Tramonto}
 \affiliation{Dipartimento di Fisica, Universit\`a degli Studi
              di Milano, via Celoria 16, 20133 Milano, Italy}
\author{D.E. Galli}
 \affiliation{Dipartimento di Fisica, Universit\`a degli Studi
              di Milano, via Celoria 16, 20133 Milano, Italy}
\author{L. Reatto}
 \affiliation{Dipartimento di Fisica, Universit\`a degli Studi
              di Milano, via Celoria 16, 20133 Milano, Italy}
\date{\today}

\begin{abstract}
We have studied a two-dimensional triangular commensurate crystal of $^4$He with the 
exact $T=0$~K Path Integral Ground State (PIGS) Monte Carlo method.
We have projected onto the true ground state both a Jastrow-Nosanow wave 
function, in which equilibrium positions are explicitly given and no 
Bose--Einstein (BEC) is present, and a translationally invariant shadow 
wave function, in which the solid phase emerges through a spontaneously 
broken symmetry process and it has BEC.
We find a remarkable convergence to the same properties, both the diagonal 
ones as well as the off--diagonal one--body density matrix $\rho_1$.
This supplies a strong evidence that no variational bias are present in the
PIGS method.
We find no BEC in the commensurate 2D $^4$He crystal at $T=0$~K, $\rho_1$ 
shows an exponential decay in the large distance range. 
The structure found in $\rho_1$ is due to virtual vacancy--interstitial
pairs and this shows up in the momentum distribution.
\end{abstract}


\maketitle

Quantum Monte Carlo methods have provided a very powerful tool in exploring the 
physics of strongly interacting many-body quantum systems. 
As far as the properties of Bose fluids and solids are concerned, Path Integral 
Monte Carlo (PIMC) methods have been proved to evaluate ``exact'' and ``unbiased''
expectation values on the thermal equilibrium state at finite 
temperature~\cite{Ceperley}; ``exact'' means that the obtained results are the true
expectation values within the statistical error, while ``unbiased'' means that the
only required input is the interatomic potential.    
At zero temperature, different ``exact'' techniques are available, as the Green's
function Monte Carlo~\cite{Kalos1} (GFMC), the Diffusion Monte Carlo~\cite{Kalos2} 
(DMC), the Reptation Monte Carlo~\cite{Baroni} or the Path Integral Ground 
State~\cite{pigs} (PIGS) methods.
All such methods, though ``exact'' in suitable limits, rely on variational models 
for the ground state wave function (wf) of the system.
These variational models play a relevant role in the importance sampling employed 
by these methods; although, in principle, the final results should not be affected
by the particular choice of the trial wf, in practice, the possibility that some bias 
could survive, especially in systems with complex broken symmetries like in the solid
phase, has not been established.
In this paper we will show the robustness of the results given by the PIGS method
with respect to the choice of the variational wf by studying a model for $^4$He in
two dimensions in the solid phase.
In fact, we find the same physical properties of the system, within the statistical
noise, starting from two radically different wfs.
One, a Jastrow--Nosanow~\cite{Nosanow} wf (JNWF), has built in the crystal lattice
via Gaussian localization factors, it is not Bose--symmetric and has no 
Bose--Einstein condensation (BEC).
The other, a shadow~\cite{Viti} wf (SWF), is translationally invariant, with 
crystalline order arising as spontaneous broken symmetry, and it has 
BEC~\cite{Masserini,Galli,Galli3}.
We have chosen to study this model in 2D for different reasons.
The reduced dimensionality allows us to study correlations to much larger distances
than in 3D, even more than 20 lattice parameters.
Fluctuations are expected to be stronger in 2D so this is a more stringent test for
convergence given the different symmetry properties of the starting variational 
wfs.
Finally, the 2D system is a model which is relevant for adsorbed $^4$He atoms on a
smooth substrate like graphite.

With the present study we address also the important question of the 
supersolid state of matter~\cite{Chan,Chan2,Reppy,Shira,Kubota,Kojima}.
PIMC computations give strong evidence that in a 3D commensurate solid $^4$He 
(number of atoms equal to the number of lattice sites) there is no superfluid 
response and no BEC~\cite{Clark,worm}.
The PIMC computations are at finite temperature and the lowest $T$ is of order 
$0.1$~K which is above the experimental transition temperature of crystals of good 
quality~\cite{Chan2}.
By computing the density matrix $\rho_1(\vec{r},\vec{r}\,')$ of crystalline $^4$He at
$T=0$~K, we find that there is no BEC in a 2D crystal. Preliminary results indicate
that this is true also in 3D.

Dealing with low temperature properties, $^4$He atoms are described as 
structureless zero--spin bosons, interacting through a realistic two--body 
potential, that we assume to be the HFDHE2 Aziz potential~\cite{Aziz}.
The aim of the PIGS method is to improve a variationally optimized trial wf by 
constructing a path in the Hilbert space of the system which connects the given wf 
to the true ground state of the system; during this ``path'', the correct 
correlations among the particles arise through the ``imaginary time evolution 
operator'' $e^{-\tau\hat H}$, where $\hat H$ is the Hamiltonian operator.
Being $\phi$ a trial wf with non--zero overlap with the exact ground state $\psi_0$, 
this $\psi_0$ is obtained as the $\tau\to\infty$ limit of
$\psi_\tau=e^{-\tau\hat H}\phi$ suitably normalized. This $\psi_\tau$ can be written analytically 
by discretizing the path in imaginary time and exploiting the factorization property 
$e^{-(\tau_1+\tau_2)\hat H} =e^{-\tau_1\hat H}e^{-\tau_2\hat H}$.
In this way, $\psi_\tau$ turns out to be expressed in term of convolution integrals 
which involve the ``imaginary time propagator'' $\langle R|e^{-\delta\tau\hat H}|R'\rangle$ 
for a $\delta\tau$, that can be small enough such that 
very accurate approximants are known~\cite{Ceperley,Suzuki}.
This maps the quantum system into a classical system of open polymers~\cite{pigs}.
An appealing feature peculiar to the PIGS method is that, in $\psi_\tau$, the 
variational ansatz acts only as a starting point, while the full path in imaginary 
time is governed by $e^{-\tau\hat H}$, which depends only on the Hamiltonian operator. 
  
As a trial wf, we used both a JNWF and a SWF.
The JNWF is written as the product of two--body correlations and of Gaussian one--body 
terms which localize the particles around the assumed lattice positions~\cite{Nosanow}.
In the SWF, beyond the explicit two--body factors, additional correlations are 
introduced via auxiliary (shadow) variables which are integrated out~\cite{Viti}.
Nowadays, SWF gives the best variational description of solid and liquid 
$^4$He~\cite{Moroni}.
In both the cases, the variational parameters have been chosen to minimize the
expectation value of the Hamiltonian operator.
In what follows, we will refer to PIGS when we deal with the projection of the 
JNWF, and to SPIGS~\cite{spigs} (Shadow Path Integral Ground State) when we project 
the SWF.

Because of the Bose statistics obeyed by the atoms, when using $\psi_\tau$ as an
approximation of the true ground state $\psi_0$, one has, in principle, also to account 
for permutations in the propagator 
$\langle R|e^{-\delta\tau\hat H}|R'\rangle $~\cite{Ceperley,pigs}.
Permutation moves are necessary when the JNWF, which is not
Bose--symmetric, is used.
On the other hand permutation moves are not necessary whenever the trial wf in $\psi_\tau$ 
is already Bose--symmetric, as the SWF. 
However, also for SPIGS, adding permutation moves turns out to be useful
in improving the efficiency and the ergodicity of the sampling, mainly in reaching the
large--distance range of $\rho_1$.
In our algorithm we have introduced two different permutation samplings: cycles of
inter--particles exchanges and swap moves.
The first, which may involve an arbitrary number of particles, are described in
detail in Ref.~\onlinecite{scambi}.
In a PIGS or SPIGS calculation of $\rho_1(\vec{r},\vec{r}\,')$, the efficiency can be further improved 
by introducing particular two--particles permutations cycles involving always one of 
the two positions $\vec{r}$ and $\vec{r}\,'$: the swap moves~\cite{worm}.
These moves improve very much the efficiency of the computation of $\rho_1$ 
and their acceptation rate is remarkably high: in the present 2D system,
by using a staging\cite{molotrovo} method to sample the free (kinetic) part of the
imaginary time propagator,
we have found an acceptation rate for this swap move which is nearly 15\%
in the liquid phase and in the solid phase at densities close to the melting.

\begin{figure}[t]
 \includegraphics[width=8cm]{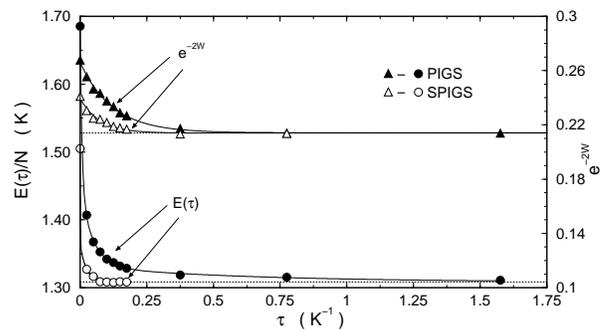}
 \caption{\label{f:over} Energy per particle $E/N$ and Debye--Waller factor $e^{-2W}$
          as functions of the projection time $\tau$ obtained from PIGS and SPIGS for a 
          commensurate 2D $^4$He crystal with $N=120$ particles at $\rho = 0.0765$~\AA$^{-2}$.
          Error bars are smaller than the used symbols.
          The dotted lines indicate the convergence values $\epsilon_0=1.308\pm0.002$~K and 
          $e^{-2W}=0.214\pm0.002$, solid lines are exponential fits to guide the eye.}
\end{figure}
The 2D $^4$He system phase diagram is known from accurate finite temperature PIMC
simulations~\cite{Gordillo}; at zero temperature both DMC~\cite{DMC} and 
GFMC~\cite{GFMC} have been used to investigate its properties mainly for the liquid 
phase.
We have performed SPIGS and PIGS simulations of a 2D $^4$He commensurate triangular crystal
at $\rho=0.0765$~\AA$^{-2}$, slightly above the melting density.
In order to control the reliability of our results we have tested their dependence 
on both the ``projection time'' $\tau$ and the ``time step'' $\delta\tau$.
For a fixed value of $\tau = 0.075$~K$^{-1}$, we have done calculations with 
$\delta\tau = 1/40$, $1/80$, $1/160$ and $1/320$~K$^{-1}$ and we have used the 
pair--product approximation~\cite{Ceperley} for the imaginary time propagator.
Reducing $\delta\tau$ below $1/40$~K$^{-1}$ affects only marginally the results;
for example, by using $1/320$~K$^{-1}$, the obtained energy is only 1\% lower then the
one with $1/40$~K$^{-1}$. So we have adopted in most of the computations 
$\delta\tau = 1/40$~K$^{-1}$ as a reasonable compromise between accuracy and 
computational effort.
These tests provide also a robust check of the ergodicity of the sampling algorithm since a lower 
value of $\delta\tau$ for given $\tau$ means a bigger number of small time projections 
(convolution integrals) so that one is dealing with polymers of increasing length.
We have then increased $\tau$ till convergence in the results has been achieved.

Diagonal properties, like the energy, have been computed in a box which hosts $N=120$
particles with periodic boundary conditions.
In Fig.~\ref{f:over} we give the energy per particle as a function of 
$\tau$ both for SPIGS and PIGS: at $\tau=0.075$~K$^{-1}$ the SPIGS result is 
already converged to the value $E/N=1.309\pm0.002$~K, while with the PIGS one reaches 
convergence at $\tau=1.575$~K$^{-1}$, where the energy takes the value 
$E/N=1.311\pm0.002$~K.
The potential energy values are respectively $-11.053\pm0.004$~K and 
$-11.048\pm0.006$~K.
From Fig.~\ref{f:over}, it is evident that the true ground state value is reached 
much more quickly (i.e. with a lower number of small-time projections) with SPIGS.
The overlap between the trial wf and the true ground state plays a crucial role in
the convergence: a large overlap ensures that $\psi_\tau$ is a 
good approximation even for not too large $\tau$.
It has been shown~\cite{overlap} that $|\langle \psi_0|\phi \rangle|^2=(e^{-\gamma})^N$ where
$\gamma$ corresponds to the area of the region between $E(\tau)/N$ and 
its convergence value $\epsilon_0$.
From our results, $e^{-\gamma}$ turns out to be 99.8\% for the SWF and 
97.9\% for the JNWF.
From the peaks in the static structure factor we have extracted the Debye-Waller factor; we have 
found that both SPIGS and PIGS converge to the same value $e^{-2W}=0.214\pm0.002$ 
(Fig.~\ref{f:over}) and, as in the energy case, SPIGS shows a faster convergence.
It is important to note that the energy convergence does not allow to deduce 
{\it a priori} the convergence of other physical properties: the convergence must 
be checked independently for each observable.

\begin{figure}[t]
 \includegraphics[width=8cm]{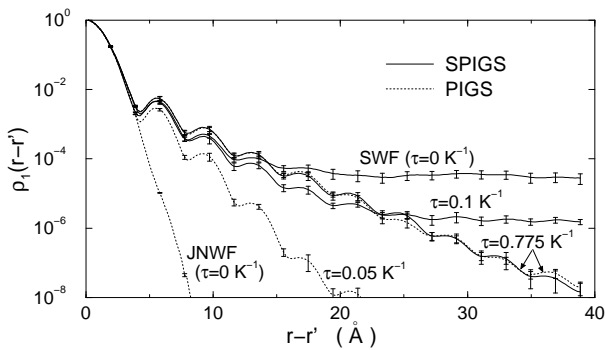}
 \caption{\label{f:con} One body density matrix $\rho_1(\vec{r}-\vec{r}\,')$
          computed along the nearest--neighbor direction in a 2D commensurate
          $^4$He crystal at the density $\rho=0.0765$~\AA$^{-2}$ for different
          $\tau$ values. Notice the completely different behavior for the two
          initial wfs and the convergence to the same $\rho_1$.}
\end{figure}
In order to study whether the 2D commensurate solid $^4$He has BEC, we have computed the
one--body density matrix $\rho_1(\vec{r},\vec{r}\,')$.
A non--zero limit of $\rho_1$ for $|\vec{r}-\vec{r}\,'|\to\infty$ implies BEC and 
the Fourier transformation of $\rho_1$ gives the momentum distribution 
$n_{\vec{k}}$.
We have computed $\rho_1$ in a 2D commensurate crystal at $\rho=0.0765$~\AA$^{-2}$
in a simulation box with $N=240$ particles (this allows us to explore distances
up to about 39~\AA).
In our calculations we have sampled $\rho_1(\vec{r},\vec{r}\,')$ in two 
different ways.
In the first one we force $\vec{r}-\vec{r}\,'$ to lie on a nearest--neighbor 
direction, while, in the second $\vec{r}$ and $\vec{r}\,'$ are allowed to explore 
the full plane; we have found perfectly consistent results.
In Fig.~\ref{f:con} we show the convergence of $\rho_1$ computed along the
nearest--neighbor direction with SPIGS and PIGS for increasing projection time.
The $\rho_1$ given by SWF has a non--zero limit at large distance, i.e. there is BEC
as in 3D~\cite{Galli}.
When projecting from SWF the large distance plateau decreases for increasing $\tau$
until, at $\tau=0.775$~K$^{-1}$, it disappears up to the distance allowed by the 
simulation box.
At convergence, $\rho_1$ displays an exponential decay, with a correlation length 
$\lambda\simeq2.75$~\AA, with superimposed a small modulation reflecting the 
crystal symmetry.
The $\rho_1$ given by the JNWF is essentially a Gaussian.
Under projection, exchanges among the atoms greatly extended the range of 
$\rho_1$.
It is remarkable that at $\tau=0.775$~K$^{-1}$ the $\rho_1$ obtained starting from
such different wfs coincide within the statistical uncertainty.
Similar conclusions are reached also from the computation of $\rho_1$ in the whole 
plane.
\begin{figure*}[t]
 \begin{center}
 \subfigure{\includegraphics[width=8.5cm,height=5cm]{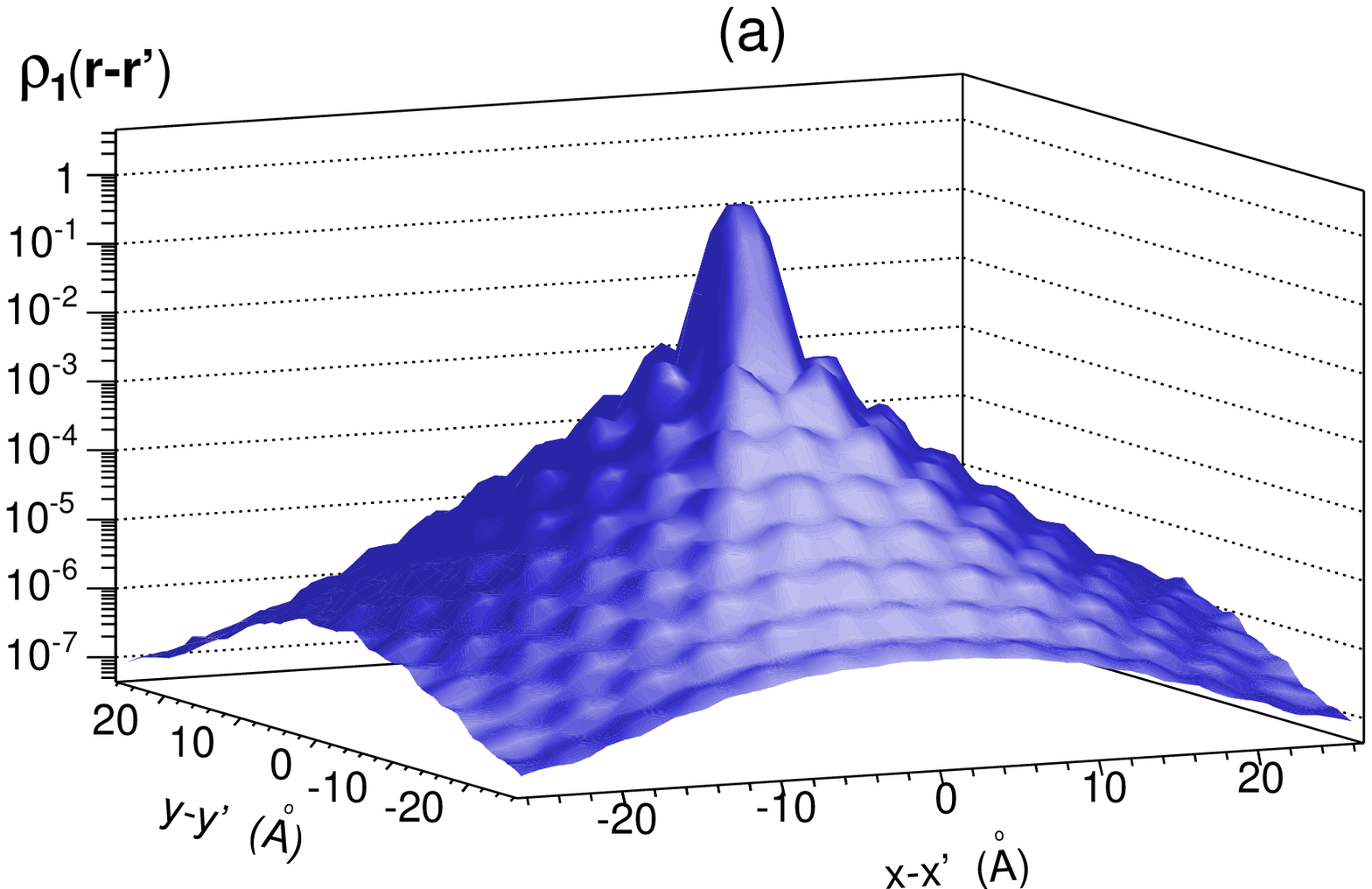}}
 \subfigure{\includegraphics[width=8.5cm,height=5cm]{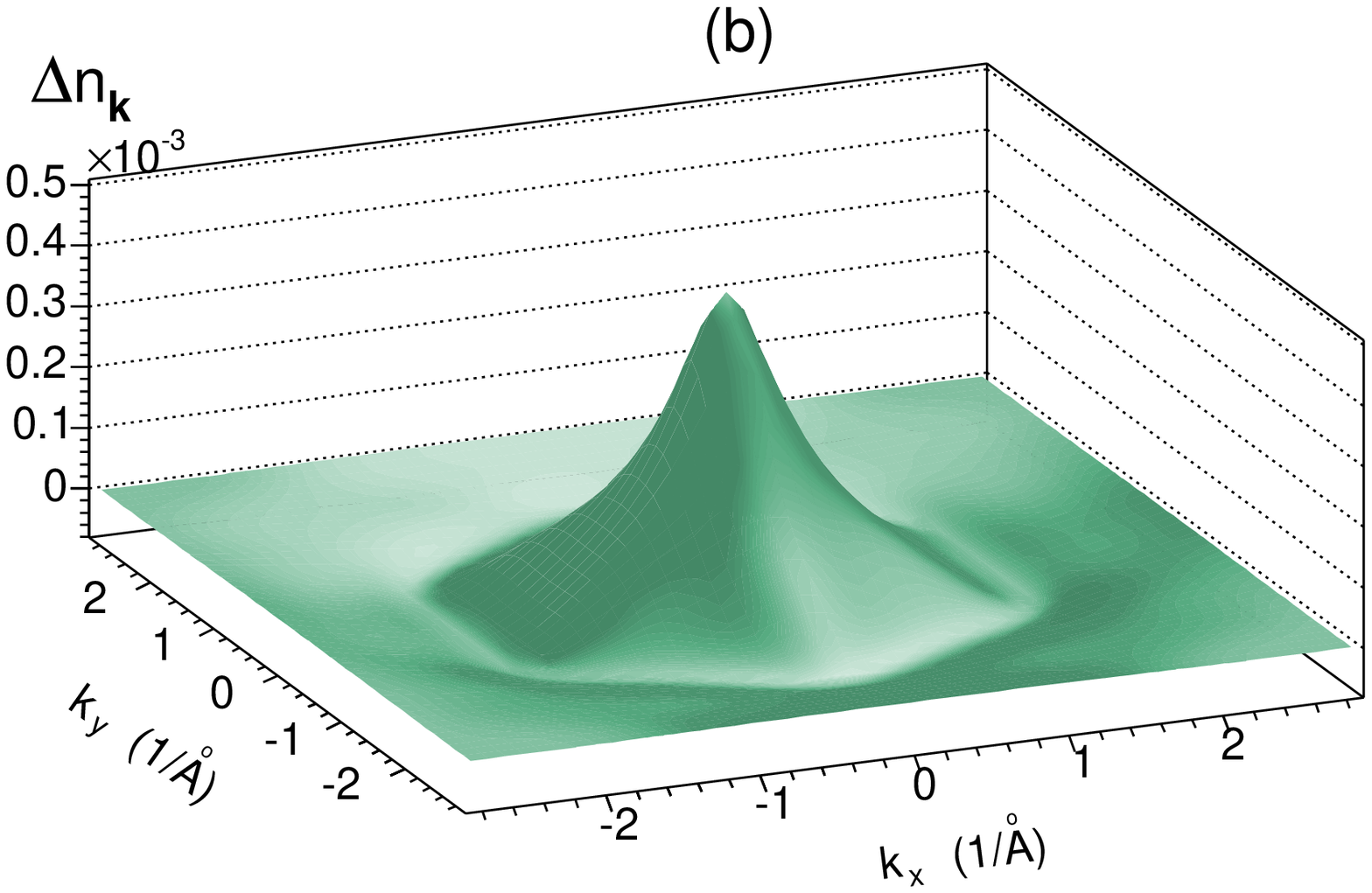}}
 \caption{\label{f:rho} (a) One--body density matrix $\rho_1(\vec{r}-\vec{r}\,')$ 
          computed in a 2D commensurate $^4$He crystal with 224 particles at the 
          density $\rho=0.0765$~\AA$^{-2}$, in the $x-x'>0$ ($x-x'<0$) range the
          SPIGS (PIGS) results are displayed.
          (b) Deviation of the momentum distribution $\Delta n_{\vec{k}}$ from
          the Gaussian distribution $n_{\vec{k}}=\frac{2\pi\hbar^2}{m\langle T\rangle}
          \exp{\left(-\frac{\hbar^2|\vec{k}|^2}{2m\langle T \rangle}\right)}$, obtained 
          by Fourier transforming the difference between the SPIGS $\rho_1$ and the Gaussian 
          distribution corresponding to the kinetic energy $\langle T \rangle$.}
 \end{center}
\end{figure*}
In Fig.~\ref{f:rho}a we show $\rho_1$ obtained both with SPIGS ($x-x'>0$ range) and 
PIGS ($x-x'<0$ range) with the projection time $\tau=0.775$~K$^{-1}$.
Again, the results are indistinguishable within the statistical error.
We have looked also for finite size effects by considering system at the same 
density but with different particles number ($N=180$, 240 and 480) and we have 
found no appreciable differences. 
With PIGS we have studied $\rho_1$ up to a distance of about 60~\AA, finding the 
same exponential decay. 

We conclude that a 2D commensurate $^4$He crystal has no BEC, or, more precisely,
the condensate fraction, if any, is below 10$^{-8}$.
We notice that exchange processes are rather significant in the system, so that the
range of $\rho_1$ is significantly larger than the size of the unit cell.
This manifests itself in the momentum distribution (Fig.~\ref{f:rho}b) 
as deviation $\Delta n_{\vec{k}}$ of $n_{\vec{k}}$ from the Gaussian corresponding 
to the kinetic energy.
There is an excess of particles at low momenta up to $k=1.2$~\AA$^{-1}$ 
whereas there is a deficit in the $k$--space region around $k=1.6$~\AA$^{-1}$.
The positions of the bumps of $\rho_1$ over the exponential decay suggest that the
extension of $\rho_1$ beyond the unit cell can be interpreted as a signal of the appearence
of vacancy-interstitial pairs (VIP): their position respect to the origin $(\vec{r}=\vec{r}\,')$
correspond to interstitial positions whereas the distance between the positions of two
neighboring pumps is equal to the lattice parameter.
In the case of the SWF these VIP are unbound allowing for a finite BEC~\cite{Galli},
whereas these pairs are bounded in the exact $\psi_0$.
This might be due to the presence of some long range correlations in $\psi_0$,
possibly caused by the zero--point motion of transverse phonons, which
are absent in the SWF.
Since the ground state wf reflects the zero--point motion of any excitation in the 
system~\cite{Reatto}, it is tempting to suggest that the evidence of VIP in $\rho_1$
is the manifestation of a new kind of excitation in the quantum solid beyond the 
phonon excitations.

Our results that the 2D commensurate crystal has no BEC at $T=0$~K and the similar
results~\cite{Clark,worm} obtained in 3D at finite $T$ make plausible that also in 3D
a commensurate crystal has no BEC at $T=0$~K and, in fact, we have some preliminary
results for $\rho_1$ in 3D supporting this notion.
Experimentally it is established that defects and $^3$He impurities have an 
important role on the non--classical rotational inertia (NCRI) of $^4$He crystal 
and dislocations are suspected to have a relevant role~\cite{Chan2}, at least in the
case of good quality crystals.
It is an open question what happens in a solid with less and less defects: will any
NCRI effect go away?
Our result suggests that it should go away unless some disorder is present even in
the ground state as an intrinsic property of $\psi_0$.
A key question is therefore to establish whether zero--point defects are present
in the ground state of a quantum solid.
We believe that the presence of zero--point defects in solid $^4$He is still an
open question~\cite{Galli2}.
Since a SWF, which has BEC, is the exact ground state of a suitable Hamiltonian, the interesting
question is for which class of interatomic potentials does the commensurate crystal
have BEC and supersolidity?

The convergence in the results obtained starting from radically different wfs is a remarkable
result because supplies strong evidence for the absence of any variational bias in 
the PIGS method.
Moreover, even if not shown in this letter, we have also obtained convergence
of diagonal properties, such as the static structure factor and the radial distribution 
function, by projecting a simple Jastrow wf which has just the minimal information 
on the short range behavior and displays no crystalline order at the considered density.

This work was supported by the INFM Parallel Computing Initiative, by the 
Supercomputing facilities of CILEA and by the Mathematics Department 
``F. Enriques'' of the Universit\`a degli Studi di Milano.

\end{document}